\input harvmac
\def\Dslash{D\!\!\!\! /}

\def\ket{\!\!>}
\def\bra{<\!\!}
\def\frak#1#2{{\textstyle{{#1}\over{#2}}}}

\def\bz{\overline z}
\def\zhat{\hat z}
\def\zhatk{\hat z^{(k)}}

\def\pa{\partial}
\def\dpl{\pa_+}
\def\dmi{\pa_-}
\def\semi{;\hfil\break}

\def\Qtilde{\tilde Q}
\def\Ocal{{\cal O}}

\def\sic{supersymmetric}
\def\DRED{\ifmmode{{\rm DRED}} \else{{DRED}} \fi}
\def\DREDp{\ifmmode{{\rm DRED}'} \else{${\rm DRED}'$} \fi}
\def\NSVZ{\ifmmode{{\rm NSVZ}} \else{{NSVZ}} \fi}

\def\npb{{Nucl.\ Phys.\ }{\bf B}}
\def\prd{{Phys.\ Rev.\ }{\bf D}}

\def\plb{{Phys.\ Lett.\ }{\bf B}}

{\nopagenumbers
\line{\hfil LTH 561}
\line{\hfil hep-th/0209111}
\vskip .5in
\centerline{\titlefont General classical solutions }
\centerline{\titlefont in the noncommutative $CP^{N-1}$ model}
\vskip 1in
\centerline{\bf O.~Foda${}^*$, I.~Jack${}^{\dagger}$ and 
D.R.T.~Jones${}^{\dagger}$}
\medskip
\centerline{\it ${}^*$Dept. of Mathematics and Statistics,
University of Melbourne,}
\centerline{\it Parkville, Victoria 3052, Australia}
\vskip 10pt
\centerline{\it ${}^{\dagger}$Dept. of Mathematical Sciences,
University of Liverpool, Liverpool L69 3BX, UK}

\vskip .3in

We give an explicit construction of general classical solutions for the
noncommutative $CP^{N-1}$ model in two dimensions, showing that they correspond 
to integer values for the action and topological charge. We also give explicit 
solutions for the Dirac equation in the background of these general solutions
and show that the index theorem is satisfied.

\Date{September 2002}}

The two-dimensional $CP^{N-1}$ model\ref\cpref{H.~Eichenherr, \npb146 (1978) 215
\semi 
E.~Cremmer and J.~Scherk, \plb74 (1978) 341}\ref\golo{V.~Golo and 
A.~Perelomov, \plb79 (1978) 112}
has long been considered as a
useful test-bed for four-dimensional gauge theories, since it possesses many of
the same features, such as conformal invariance, asymptotic freedom and 
a topological charge taking integer values. There is also the 
possibility of a ${1\over{N}}$ expansion giving useful information about
confinement\ref\dadda{A.~D'Adda, P.~di Vecchia and M.~L\"uscher, 
\npb146 (1978) 63}\ref\daddb{A.~D'Adda, P.~di Vecchia and M.~L\"uscher, 
\npb152 (1978) 125}. The action is minimised for 
self-dual and anti-self-dual instanton configurations\golo\dadda\ 
for which the action
is proportional to the topological charge. The instanton contribution
to the functional integral defining the quantum field theory was 
evaluated in Refs.~\ref\dlz{
A.M.~Din, P.~di Vecchia and W.J.~Zakrzewski, \npb155 (1979) 
447}\ref\bl{B.~Berg and M.~L\"uscher, Comm. Math. Phys. 69 
(1980) 57}. However, the instanton and 
anti-instanton solutions do not exhaust the solutions to the classical 
field equations which should be used in this stationary phase 
approximation; it turns out that the general solution can be 
given by a very elegant 
construction\ref\dza{A.M.~Din and W.J.~Zakrzewski, \npb174 (1980) 
397}\ based on a similar analysis for the $O(2k+1)$ $\sigma$-model in
Ref.~\ref\borch{H.J.~Borchers and W.D.~Garber, Comm. Math. Phys. 72 (1980) 77}.
These solutions are in general saddle points of the action, which still
takes well-defined integer values as of course does the topological 
charge\ref\dzb{A.M.~Din and W.J.~Zakrzewski, \plb95 (1980) 419}. Moreover the
solution to the Dirac equation in the background of a general classical 
solution can be given in an equally elegant fashion\ref\dzc{A.M.~Din and 
W.J.~Zakrzewski, \plb101 (1981) 166}.

Quantum field theory on noncommutative spacetime has received much attention
recently,
largely due to its emergence in $M$-theory (for reviews see 
Refs.~\ref\DouglasBA{ M.~R.~Douglas and N.~A.~Nekrasov, 
hep-th/0106048}\ref\szabo{R.J.~Szabo, hep-th/0109162}). In 
particular a good deal 
of work has been devoted to instanton solutions in noncommutative gauge 
theory\ref\inst{N.A.~Nekrasov and A.~Schwarz, Comm. Math. Phys. 198
(1998) 689 [hep-th/9802068]\semi
K.~Lee and P.~Yi, \prd61 (2000) 125015 [hep-th/9911186]\semi
S.~Terashima, \plb477 (2000) 292 [hep-th/9911245]\semi
M.~Marino, R.~Minasian, G.~Moore and A.~Strominger, JHEP 0001 
(2000) 005 [hep-th/9911206]\semi
K.~Furuuchi, Prog. Theor. Phys. {\bf 103} (2000) 1043 [hep-th/9912047]\semi
K.-Y.~Kim, B.-H.~Lee and H.S.~Yang, hep-th/0003093; \plb523 (2001) 357
[hep-th/0109121];
\prd66 (2002) 025034
 [hep-th/0205010]  \semi
D.J.~Gross and N.A.~Nekrasov, JHEP 0007 (2000) 034 [hep-th/0005204]\semi
M.~Hamanaka,
Phys.\ Rev.\ {\bf D}65 (2002) 085022
[hep-th/0109070]\semi
Y.~Tian and C.-J.~Zhu, hep-th/0205110\semi
B.-H.~Lee and H.S.~Yang, \prd66 (2002) 045027 [hep-th/0206001]}. As in 
the commutative case, this has also naturally led to interest in the $CP^{N-1}$ 
model\ref\LeeEY{B.H.~Lee, K.M.~Lee and H.S.~Yang, 
Phys.\ Lett.\ {\bf B}498 (2001) 277 [hep-th/0007140]}. 
It was shown that the instanton and 
anti-instanton solutions could straightforwardly be generalised to the 
noncommutative case. In this paper we shall show that all the work on the 
general classical solutions in Refs.~\dza--\dzc\ generalises equally 
naturally. 

We consider the $CP^{N-1}$ model defined on the noncommutative complex 
plane\LeeEY. 
The co-ordinates $x_1$, $x_2$ satisfy
\eqn\comm{
[x_1,x_2]=  i\theta,}
with $\theta>0$, 
but as usual it will prove useful to introduce complex co-ordinates
\eqn\compdef{
x_+=  {x_1+ix_2\over{\sqrt2}},\qquad x_-=  {x_1-ix_2\over{\sqrt2}}}
which satisfy
\eqn\newcomm{
[x_+,x_-]=  \theta.}
We can represent $x_{\pm}$ by creation and annihilation 
operators, acting on the harmonic oscillator Hilbert space with ground state
$|0\ket$ satisfying $x_+|0\ket= 0$ and 
$|n\ket= {1\over{\sqrt{\theta^nn!}}}(x_-)^n|0\ket$, 
so that
\eqn\updown{
x_+|n\ket= \sqrt{\theta n}|n-1\ket,\quad x_-|n\ket= \sqrt{\theta(n+1)}|n+1\ket.}
Then the derivatives $\pa_{\pm}= {\pa\over{\pa x_{\pm}}}$ can be 
represented as 
\eqn\dascom{
\pa_+z= -\theta^{-1}[x_-,z],\quad \pa_-z= \theta^{-1}[x_+,z].}
Correspondingly, the integration over the noncommutative plane 
may be represented by the trace over the harmonic oscillator Hilbert space. 
The Lagrangian is given by
\eqn\laga{L=  \pa_{\mu}\bz.\pa^{\mu}z+(\bz.\pa_{\mu}z)(\bz.\pa^{\mu}z)
=  \overline{D_{\mu}z}.D_{\mu}z,}
where $z$ is an $N$-dimensional vector over the noncommutative complex plane,
subject to the constraint
\eqn\modsq{
\bz. z=  1,}
where for any $N$-vector $X$
\eqn\covdef{
D_{\mu}X=  \pa_{\mu}X-X\bz.\pa_{\mu}z,}
and 
\eqn\moddef{|X|^2 =  \overline{X}.X}
(Strictly speaking it is an abuse of terminology to call $z$ a vector since 
vector spaces are defined over a field, in which by definition multiplication is
commutative. However, all the standard theorems and properties of 
vector spaces with regard to bases, dimensionality and orthogonality will
remain valid in the noncommutative case, as long as we adhere to the convention 
that multiplication of a vector by a scalar is on the right.) 
Note that the complex conjugate satisfies $\overline{fg}= \overline{g}
\overline{f}$.

The corresponding action is 
\eqn\acta{
S=  \Tr L=  2\pi\theta\sum_{n\ge0}\bra n|L|n\ket,}
where $|n\ket$, $n=  0,1,2\ldots$ are the usual normalised harmonic oscillator 
energy eigenstates. 
The model has a local $U(1)$ symmetry 
\eqn\symm{
z\rightarrow zg(x),\quad g(x)\in U(1),}
under which 
\eqn\Xtran{
D_{\mu}X\rightarrow D_{\mu}Xg(x).}
Note that here as always the ordering is crucial. On introducing 
covariant derivatives $D_{\pm}$ corresponding to $x_{\pm}$, we can rewrite 
$L$ as 
\eqn\lrewrite{
L=  |D_{+}z|^2+|D_{-}z|^2,}
and the field equation can be written in the equivalent forms
\eqn\field{\eqalign{
D_+D_-z+z|D_{-}z|^2=&  0,\cr
\hbox{or alternatively}\quad D_-D_+z+z|D_{+}z|^2=&  0.\cr}}
The commutative $CP^{N-1}$ model has well-known instanton and anti-instanton 
solutions satisfying 
\eqn\inst{
D_{\pm}z=  0,}
given by
\eqn\insta{
z=  f(x_{\mp}){1\over{|f(x_{\mp})|}}.}
Here due to gauge invariance $f$ may be assumed without loss of generality
to be a polynomial $N$-vector.
It can be shown that (with this ordering) these remain solutions in the 
noncommutative case. A feature which appears here and elsewhere is that 
awkward derivatives such as that of ${1\over{|f(x_{\mp})|}}$ get projected out.

It was shown in Ref.~\dza\ that the commutative $CP^{N-1}$ model has additional 
classical solutions which are neither instantons nor anti-instantons, and these 
were presented explicitly using a simple, elegant construction. We shall show 
here that these solutions (correctly ordered) remain valid in the 
noncommutative case. We claim that a general solution is given by
\eqn\sola{
z^{(k)}=  \zhatk{1\over{|\zhatk|}},}
where
\eqn\solb{
\zhatk=  \pa^k_+f-\sum_{l,m=  0}^{k-1}\dpl^lf M^{-1}_{l,m}\dpl M_{m,k-1},}
where $f(x_+)$ is a polynomial $N$-vector\foot{As noted in Ref.~\LeeEY,
the use of functions  with singularities presents difficulties in the
noncommutative case} and the matrix $M$ has entries
\eqn\Mdef{
M_{l,m}=  \overline{\dpl^lf}.\dpl^mf, \qquad l,m=  0,\ldots k-1.}
(We assume that $f,\dpl f,\ldots\dpl^{N-1}f$ are linearly independent.) 
We start by noting the following identities which follow immediately from 
Eq.~\solb: 
\eqna\dminus$$\eqalignno{
\overline{\zhatk}.\dmi\zhatk=  &0,&\dminus a\cr
\overline{\dpl^if}.\zhatk=  & \delta^{ik}|\zhatk|^2,
\quad i= 0,1,\ldots k,&\dminus b\cr
\overline{\zhatk}.\pa_+\zhat^{(i-1)}=&\delta^{ik}|\zhatk|^2,
\quad i= 0,1,\ldots k.&\dminus c\cr}$$
It is then easy to establish the following results, using Eq.~\dminus{a}:
\eqna\lemmasa$$\eqalignno{
D_+X= &\dpl\left(X{1\over{|\zhatk|}}\right)|\zhatk|,&\lemmasa a\cr
D_-X= &\dmi\left(X|\zhatk|\right){1\over{|\zhatk|}}.&\lemmasa b\cr}$$
Two additional useful identities are as follows:
\eqna\lemmas$$\eqalignno{
\dpl\left(\zhat^{(k)}{1\over{|\zhat^{(k)}|^2}}\right) &=  \zhat^{(k+1)}
{1\over{|\zhat^{(k)}|^2}},&\lemmas a\cr
\dmi\zhat^{(k+1)} &=  -\zhat^{(k)}{1\over{|\zhat^{(k)}|^2}}|\zhat^{(k+1)}|^2.
&\lemmas b\cr}$$
To prove Eqs.~\lemmas{}, we start by defining 
\eqn\lspace{
L_i=  \{f,\dpl f,\ldots\dpl^{i-1}f\}= \{\zhat^{(0)},\ldots\zhat^{(i-1)}\},}
where by this notation we mean that $L_i$ is the subspace whose basis is as
shown
(the second equality holds since it is clear from Eqs.~\solb, \dminus{b}\ that
$\overline{\zhat^{(i)}}.\zhat^{(i')}=  0$ for $i\ne i'$).  
Then Eqs.~\lemmas{}\
are easily proved, by first noting that the LHS of Eq.~\lemmas{a}\ is 
clearly in $L_{k+2}$ and that of Eq.~\lemmas{b}\ is clearly in $L_{k+1}$.
Then expanding the LHS of each identity in terms of the basis vectors
$\zhat^{(i)}$, we may readily establish the coefficients. For 
instance, we may write (using Eq.~\dminus{a})
\eqn\zexp{
\dpl\left(\zhat^{(k)}{1\over{|\zhat^{(k)}|^2}}\right)= \sum_{j= 0}^{k+1}
\zhat^{(j)}\alpha^{(j)}= 
\dpl\zhatk{1\over{|\zhatk|^2}}-\zhatk{1\over{|\zhatk|^2}}\overline{\zhatk}.
\dpl\zhatk{1\over{|\zhatk|^2}},}
and then take the scalar product with $\zhat^{(i)}$, $i=0,\ldots k+1$ in turn.
Then for $i= 0,1,\ldots k-1$, 
\eqn\lemmasb{|\zhat^{(i)}|^2\alpha^{(i)}= 
\overline{\zhat^{(i)}}.\dpl\zhatk{1\over{|\zhatk|^2}}
= \left[\dpl(\overline{\zhat^{(i)}}.\zhatk)-\overline{\dmi\zhat^{(i)}}.\zhatk
\right]{1\over{|\zhatk|^2}}= 0}
(since $\dmi\zhat^{(i)}\in L_i$).
For $i= k$, $|\zhatk|^2\alpha^{(k)}= 0$
follows immediately from Eq.~\zexp. Finally, using Eq.~\dminus{c}\ 
we find $\alpha^{(k+1)}= {1\over{|\zhat^{(k)}|^2}}$, thus
completing the proof of Eq.~\lemmas{a}. Eq.~\lemmas{b}\ is proved in similar 
fashion.

We can now write, using Eqs.~\sola, \lemmasa{}, \lemmas{},
\eqn\solproof{\eqalign{
D_+D_-z^{(k)}= &\dpl\left(\dmi\zhatk{1\over{|\zhatk|^2}}\right)|\zhatk|\cr
= &-z^{(k)}|\zhatk|{1\over{|\zhat^{(k-1)}|^2}}|\zhatk|\cr
= &-z^{(k)}|D_-z^{(k)}|^2,\cr}}
showing that $z^{(k)}$ is a solution of Eq.~\field. 
A useful representation of these solutions\dzb\ is derived by defining the 
operator $P_+$ as
\eqn\Pplus{
P_+g= \dpl g-g{1\over{|g|^2}}(\overline g.\dpl g).}
We then have
\eqn\Back{
\zhatk= P_+^kf.}
Eq.~\Back\ is easily proved using induction. Assuming it true for 
$k= 0,1,\ldots l$, we have
\eqn\inducta{
P^{l+1}_+f= \dpl\zhat^{(l)}-\zhat^{(l)}{1\over{|\zhat^{(l)}|^2}}
\overline{\zhat^{(l)}}.\dpl\zhat^{(l)}.}
Since $\zhat^{(l)}\in L_{l+1}$, $P^{l+1}_+f\in L_{l+2}$. Since $\zhat^{(l)}$ 
is orthogonal to $L_{l}$, we have for $i\le l-1$
\eqn\inductb{
\overline{\dpl^if}.P^{l+1}_+f= \dpl\left[\overline{\dpl^if}.\zhat^{(l)}
\right]= 0,}
and also
\eqn\inductc{\eqalign{
\overline{\dpl^lf}.P^{l+1}_+f= &\dpl[\overline{\dpl^lf}.\zhat^{(l)}]
-\overline{\dpl^lf}.\zhat^{(l)}{1\over{|\zhat^{(l)}|^2}}
\overline{\zhat^{(l)}}.\dpl\zhat^{(l)}\cr
= &\left[\dpl(|\zhat^{(l)}|^2)-\overline{\zhat^{(l)}}.\dpl\zhat^{(l)}
\right]= 0,\cr}}
using Eq.~\dminus{}. Hence
$P^{l+1}_+f$ is in $L_{l+2}$ and orthogonal to $L_{l+1}$. So we must have 
$P^{l+1}_+f= \zhat^{(l+1)}\mu$ for some $\mu$, and using Eq.~\dminus{c}\
we find $\mu= 1$. Moreover the result is 
trivially true for $k= 0$, completing the inductive proof. 

We shall now show that the topological charge is given in the same way as for 
the commutative case. The action $S^{(k)}$ corresponding to a solution $z^{(k)}$
may be written 
\eqn\actb{
S^{(k)}= 2\pi\Qtilde^{(k)}+2I^{(k)},}
where the topological charge $\Qtilde^{(k)}$ is given by 
\eqn\qden{
\Qtilde^{(k)}= {1\over{2\pi}}\Tr[Q^{(k)}],} 
with the topological charge density $Q^{(k)}$ defined as
\eqn\qdena{
Q^{(k)}= |D_+z^{(k)}|^2-|D_-z^{(k)}|^2,}
and where
\eqn\Idefa{
I^{(k)}= \Tr|D_-z^{(k)}|^2.}
It can easily be shown using Eqs.~\dminus{a}, \lemmasa{}\ that
\eqn\Qres{
Q^{(k)}= |\zhatk|\dmi\Ocal^{(i)}{1\over{|\zhatk|}},}
where
\eqn\Ocaldef{
\Ocal^{(i)}= {1\over{|\zhat^{(i)}|^2}}\dpl|\zhat^{(i)}|^2,} 
and moreover (as can be established using induction combined with Eq.~\lemmas)
\eqn\Ires{
|D_-z^{(k)}|^2= |\zhatk|\dmi\sum_{i= 0}^{k-1}\Ocal^{(i)}{1\over{|\zhatk|}}.}
Inside the traces in Eqs.~\qden, \Idefa, the factors of $|\zhatk|$ and
${1\over{|\zhatk|}}$ cancel. We therefore find ourselves interested in
computing 
\eqn\Qbit{
X^{(i)}= \Tr[\dmi\Ocal^{(i)}]= 2\pi\theta
\sum_{n= 0}^{\infty}\bra n|\dmi\Ocal^{(i)}|n\ket.}
Following Ref.~\LeeEY, and using Eqs.~\updown, \dascom, we write
\eqn\Qbita{\eqalign{
X^{(i)}= 2\pi
&\sum_{n= 0}^{\infty}\left[\bra n+1|\Ocal^{(i)} x_+|n+1\ket
-\bra n|\Ocal^{(i)} x_+|n\ket\right]\cr
= &2\pi\lim_{N\rightarrow\infty}\bra N|\Ocal^{(i)} x_+|N\ket= -2\pi\theta^{-1}
\lim_{N\rightarrow\infty}\bra N|{1\over{|\zhat^{(i)}|^2}}
[x_-,|\zhat^{(i)}|^2]x_+|N\ket.\cr}}
After some use of Eq.~\newcomm, together with\LeeEY
\eqn\comid{\eqalign{
x_+g(x_-x_+)= &g(x_-x_++\theta)x_+,\cr
x_-g(x_-x_+)= &g(x_-x_+-\theta)x_-,\cr}}
we can write 
$|\zhat^{(i)}|^2= h^{(i)}(x_-x_+,\theta)$,
where $h^{(i)}(x_-x_+,\theta)$ is a homogeneous rational polynomial 
in $x_-x_+$ and $\theta$. We find
\eqn\Qbitb{\eqalign{
X^{(i)}= &-2\pi\theta^{-1}\lim_{N\rightarrow\infty}\bra N|{1\over{h^{(i)}
(x_-x_+,\theta)}}
[h^{(i)}(x_-x_+-\theta,\theta)-h^{(i)}(x_-x_+,\theta)]x_-x_+|N\ket\cr
= &-2\pi\theta^{-1}\lim_{N\rightarrow\infty}{1\over{h^{(i)}(\theta N,\theta)}}
[h^{(i)}(\theta N-\theta,\theta)-h^{(i)}(\theta N,\theta)]\theta N\cr
= &2\pi\lim_{x\rightarrow\infty}{xH^{(i)\prime}(x)\over{H^{(i)}(x)}},\cr}}
where $H^{(i)}(x)= h^{(i)}(x,0)$, corresponding to the commutative result for 
$|\zhat^{(i)}|^2$. In other words 
\eqn\poleres{
X^{(i)}= 2\pi\gamma^{(i)}}
where the commutative $|\zhat^{(i)}|^2\sim (x_-x_+)^{\gamma^{(i)}}$ for large 
$x_-$, $x_+$. It is shown in Ref.~\dzb\ that for the case where $f$ is a 
polynomial with degree $\beta$ we have 
\eqn\gamres{
\gamma^{(i)}= \beta-2i,}
leading to 
\eqn\topres{\eqalign{
\Qtilde^{(k)}= &\beta-2k\cr
I^{(k)}= &2\pi k(\beta-k+1)\cr
S^{(k)}= &2\pi[(2k+1)\beta-2k^2].\cr}}
As advertised, these are precisely the same results as obtained in the 
commutative case\dzb. 

We now discuss the generalisation of these solutions to the \sic\
$CP^{N-1}$ model, with Lagrangian
\eqn\susylag{\eqalign{
L &=  \overline{D_{\mu}z}D_{\mu}z-i\overline{\psi}\Dslash\psi\cr
&+\frak14\left[(\overline{\psi}\psi)^2+(\overline{\psi}\gamma_5\psi)^2
-(\overline{\psi}\gamma_{\mu}\psi)^2\right],\cr}}
where the fermion field is subject to 
\eqn\fermcon{
\bz.\psi=  0.} 
The solution of the full set of coupled equations for 
$z$ and $\psi$ was discussed in Ref.~\ref\dzd{A.M.~Din and W.J.~Zakrzewski, 
\npb194 (1982) 157}, using superfields. There seems no obstacle in 
principle to generalising these solutions to the noncommutative case, but the 
formalism is somewhat complex. Here instead we consider the simpler problem 
of a fermion in the 
fixed background of a bosonic solution, as in Ref.~\dzc. This can be 
considered\dzd\
as the first-order term in a Grassmann expansion of the full solution.
The Dirac equation becomes \eqn\dirac{
\Dslash\psi- z\left(\bz.\Dslash\psi\right)=  0,}
with the constraint Eq.~\fermcon. Decomposing $\psi$ in terms of eigenstates
of $\gamma_5$, 
\eqn\decomp{\psi=  \psi^+\pmatrix{1\cr-i\cr}+\psi^-\pmatrix{1\cr i\cr},}
we have
\eqn\dirpm{
D_{\pm}\psi^{\pm}=  z\lambda_{\pm},\quad \bz.\psi^{\pm}=  0,}
where $\lambda_{\pm}$ are functions of $x_{\pm}$. 

It is now easy to show using Eqs.~\lemmasa{a}, \lemmas{}\ 
that a positive helicity solution to Eq.~\dirpm\ is given by 
\eqn\poshel{
\psi^{(k)+}=  \sum_{i\ne k}\zhat^{(i)}{1\over{|\zhat^{(i)}|^2}}
\overline{\zhat^{(i)}}.g^+(x_-)|\zhat^{(k)}|,}
(where following Ref.~\dzc\ we take $g^+$ to be a polynomial) 
provided $\overline{\zhat^{(k+1)}}.g^+$ is a function of $x_-$ alone and hence
is also polynomial (any denominator is inevitably a function of $x_-x_+$). 
The form of the solution can be further restricted by requiring it to be
normalizable on a sphere. Here we face the problem of defining the normalisation
condition in the noncommutative case; a symmetric possibility is to impose
\eqn\fincond{
\Tr\left[{1\over{1+\frak{1}{2}\{x_+,x_-\}}}|\psi^{\pm}|^2\right]<\infty.}
(In fact any ordering will lead to the same conclusions.) Following a similar
procedure as in the discussion
of topological charge, we can write $|\psi^{(k)+}|^2$ as a homogeneous
function of $x_-x_+$ and $\theta$. We then find
\eqn\normpsia{
\Tr\left[{1\over{1+\frak{1}{2}\{x_+,x_-\}}}|\psi^{(k)+}|^2\right]
= \sum_{n= 0}^{\infty}{1\over{1+n+\frak12\theta}}
\sum_{i\ne k}G^{(i)}(n,\theta),}
where (since $\overline{\zhat^{(k+1)}}.g^+$ is polynomial)
$G^{(k+1)}(n,\theta)=O(n^{\Qtilde^{(k)}-\Qtilde^{(k+1)}+D})$,
where $D$ is the degree of $\overline{\zhat^{(k+1)}}.g^+$. 
Since $\Qtilde^{k}>\Qtilde^{k+1}$, for normalisability (i.e. 
convergence of Eq.~\normpsia) we must have 
$\overline{\zhat^{(k+1)}}.g^+= 0$. Repeating the argument, we deduce in turn 
that  
$\overline{\zhat^{(i)}}.g^+$ is polynomial and thence zero for 
$i= k+2,\ldots N-1$. We therefore find the 
general normalisable solution to be
\eqn\poshela{
\psi^{(k)+}=  \sum_{i= 0}^{k-1}\zhat^{(i)}{1\over{|\zhat^{(i)}|^2}}
\overline{\zhat^{(i)}}.g^+(x_-)|\zhat^{(k)}|.}

Similarly a general negative helicity solution is given by 
\eqn\neghel{
\psi^{(k)-}=   \sum_{i\ne k}\zhat^{(i)}{1\over{|\zhat^{(i)}|^2}}
\overline{\zhat^{(i)}}.g^-(x_+){1\over{|\zhat^{(k)}|}},}
where now ${1\over{|\zhat^{(k-1)}|^2}}      
\overline{\zhat^{(k-1)}}.g^-(x_+)$ is a function of $x_+$ alone and hence
polynomial. 
Proceeding as for $\psi^{(k)+}$, we find the general normalisable solution is
\eqn\neghela{
\psi^{(k)-}=   \sum_{i= k+1}^{N-1}\zhat^{(i)}{1\over{|\zhat^{(i)}|^2}}
\overline{\zhat^{(i)}}.g^-(x_+){1\over{|\zhat^{(k)}|}}.}
The counting of the number $\xi_{\pm}$ of independent solutions $\psi^{\pm}$ 
then proceeds just as in the commutative case\dzc\ with the result that 
$\xi_{\pm}$ satisfy the index theorem  
\eqn\indexth{
\xi_{+}-\xi_{-}= -NQ^{(k)}.}

We have seen that most of the results of Refs.~\dza--\dzc\ are unaffected by
the generalisation to the noncommutative case. However, it still remains to 
establish whether Eqs.~\sola, \solb\ exhaust the set of classical solutions
in the noncommutative as well as in the commutative\dza\ case. 
The answer may have to await an extension of complex 
analysis to the noncommutative context. Further work could include the 
construction of Green functions as was done in the commutative case in 
Ref.~\ref\jo{I.~Jack and H.~Osborn, J. Phys.~{\bf A}15 (1982) 245}.
Moreover, it seems likely that the solutions found for Grassmannian 
$\sigma$-models (where $z$ becomes an $N\times p$ matrix and the model has
a local $U(p)$ invariance) 
in Refs.~\ref\grass{A.M.~Din and W.J.~Zakrzewski, Lett. 
Math. Phys. 5 (1981) 553; {\it ibid}, 7 (1983) 505; \npb237 (1984) 461} will 
straightforwardly extend to the noncommutative case; indeed this has already
been shown for the pure instanton case in Refs.~\ref\ChoRB{
O.~Lechtenfeld and A.~D.~Popov,
JHEP 0111 (2001) 040 [hep-th/0106213]; 
Phys.\ Lett.\ {\bf B}523 (2001) 178 [hep-th/0108118]\semi
J.~H.~Cho, P.~Oh and J.~H.~Park,
Phys.\ Rev.\ {\bf D}66, 025022 (2002) [hep-th/0202106]}.
\vskip 12pt
\centerline{{\bf Acknowledgements}}
\vskip 5pt
IJ thanks the Dept. of Mathematics and Statistics at the University 
of Melbourne for its hospitality, and the Australian Research 
Council for financial support, while part of this work was 
carried out. DRTJ was supported by a PPARC Senior Fellowship. 

\listrefs

\end